\documentclass[wrr]{agu2001}
\newcommand{\W}{7.5cm}
\usepackage{graphicx,amsmath,amssymb}

\authorrunninghead{Boschan et al.}
\titlerunninghead{Miscible fluid displacements  inside rough fractures}

\authoraddr{}
\begin{document}
\title{Miscible displacement fronts of shear thinning fluids inside rough fractures.}

\authors{A. Boschan \altaffilmark{1,2}, H. Auradou, \altaffilmark{1}
I. Ippolito, \altaffilmark{2}
R. Chertcoff, \altaffilmark{2}
and J.P. Hulin \altaffilmark{1}}
\altaffiltext{1}
{Laboratoire Fluides, Automatique et Syst{\`e}mes Thermiques,
UMR No. 7608, CNRS, Universit{\'e} Paris 6 and 11, B{\^a}timent 502,
Universit{\'e} Paris Sud, 91405 Orsay Cedex, France.}
\altaffiltext{2}
{Grupo de Medios Porosos, Facultad de Ingenieri\'a, Universidad de Buenos Aires, Paseo Col\'on 850, 1063 Buenos-Aires, Argentina.}

\begin{abstract}
The miscible displacement of a shear-thinning fluid by another of same rheological properties is studied experimentally in a transparent fracture by an optical technique imaging
relative concentration distributions. The fracture walls have complementary  self-affine geometries and are shifted laterally in the direction perpendicular to the
mean flow velocity {\bf U} : the flow field is strongly channelized and  macro dispersion controls the front structure for P\'eclet numbers above a few units.  The global front width increases then linearly with time and reflects the velocity distribution between the different channels.  In contrast, at the local  scale, front spreading is similar to  Taylor dispersion between plane parallel surfaces. Both dispersion mechanisms depend strongly on the fluid rheology which shifts  from Newtonian to  shear-thinning when the flow rate increases. In the latter domain, increasing the concentration enhances the global front width but reduces both Taylor dispersion (due to the flattening of the velocity  profile in the gap of the fracture) and the size of medium scale front structures. 

\end{abstract}

\begin{article}
\section{Introduction}
The transport of dissolved species  in fractured
formations is of primary importance in a large number of groundwater
systems : predicting the migration rate and the 
dispersion of contaminants from a source inside or at the surface
of  a fractured rock is then relevant to many fields such as waste storage
and water management (\cite{nas,adler99,berkowitz02}).  

In the present work, these phenomena are studied experimentally by analyzing relative concentration distributions during the displacement of a transparent fluid by a miscible dyed one inside a model transparent rough fracture. 
A key characteristic of these  fractures is the self-affine geometry of their wall surfaces : it  reproduces the multiscale geometrical characteristics of many  faults and ``fresh'' fractures. For such surfaces, the variance $\Delta h^2=< (h(\vec{r}+\vec{\Delta r}) - h(\vec{r}))^2 >$ of the local height  $h(x,y)$ of the surface with respect to a reference plane verifies : 
\begin{equation}
\frac{\Delta h}{l} = (\frac{\Delta r}{l})^\zeta,
\end{equation}
in which $(x,y)$ are coordinates in the plane of the fracture,  $\zeta$ is the self-affine exponent, $l$ the topothesy (i.e.  the length scale at which the slope $\Delta h/\Delta r$ is of the order of 1). 

In this work, the rough fracture walls  have complementary geometries : they are separated by a small distance normal to their mean plane and shifted laterally relative to each other. This shear displacement induces local aperture variations : experimental and numerical investigations demonstrate that, in this case, preferential flow paths dominantly perpendicular to the shear appear~(\cite{GentierLAR97,YeoDZ98,Auradou05}). These paths strongly influence fluid transport~(\cite{Neretnieks82,Tsang87,Brown98,Becker00}), particularly when the mean flow is, as here, parallel to these channels. 

The objective of the present paper is to study the influence of the structures of the aperture field on the displacement front  of a transparent fluid by a dyed miscible one.
For that  purpose,  the displacement process is studied at different length scales in order to identify  the different front spreading mechanisms.  Practically, the displacement front is analyzed in regions of interest of variable widths $W$ perpendicular to the flow. If $W$ is smaller than the local fracture aperture,  the front spreading will be  dispersive and controlled by local mechanisms;  for large $W$ values of the order of transverse size of the fracture ($100$ times  the aperture) the front structure is controlled, on the contrary, by preferential flow paths. Additional informations on these mechanisms will be obtained from the influence of the flow velocity and of the rheology of the fluids. 
\section{Dispersion and front spreading  in rough fractures}
Previous studies by~\cite{Ippolito93,Roux98,adler99,Detwiler00} described  mixing in fractures  by a Gaussian convection-dispersion equation. They suggested that the dispersion coefficient is the sum of the contributions of geometrical and Taylor dispersion. 
The latter results from the local advection velocity gradient between the walls of the fracture : its influence is balanced by molecular diffusion across the gap. There results  a  macroscopic Fickian dispersion parallel to the flow (\cite{Taylor51,Aris58}) characterized by the coefficient : 
\begin{equation}
\frac{D}{D_m} = \tau + f\, Pe^2,
\label{eq1}
\end{equation} 
where $D_m$ is the molecular diffusion coefficient,  the P\'eclet number $Pe = Ua_0/D_m$ represents the relative influence of the velocity gradients and molecular diffusion, $U$ the average flow velocity in the whole fracture, $a$ the gap thickness, $\tau$ the tortuosity of the void space reducing the rate of longitudinal molecular diffusion~(\cite{Bear72,Drazer02}); for flat parallel plates and a Newtonian fluid, $f=1/210$.

Geometrical dispersion reflects the disorder of the velocity field in the fracture plane and may be significant  in rough fractures.  
Theoretical investigations  by~\cite{Roux98} suggest that this geometrical mechanism is important  at intermediate  $Pe$ values; at lower (resp. higher) P\'eclet numbers, molecular diffusion (resp. Taylor dispersion) are dominant. 
Scaling arguments allow in addition to estimate P\'eclet numbers corresponding to the limits of these domains : they depend on the mean, the variance and the 
correlation length of the aperture field. Such predictions are supported by experimental 
investigations on model fractures  with a relatively weak disorder of the aperture field~(\cite{Ippolito93,Ippolito94,adler99,Detwiler00}).

In natural fractures, however, experimental~(\cite{Neretnieks82,Brown98,Becker00}) and numerical~(\cite{Drazer04}) studies indicate that mass transport is strongly influenced  by large scale preferential flow channels parallel to the mean velocity.  Anomalous dispersion is then expected as  in the analogous case of porous media with strata of different permeabilities~(\cite{Matheron80}). As pointed out by ~(\cite{Roux98}), similar effects are expected in fracture if the flow velocities vary more slowly along a streamline than perpendicular to it.  Then, large distorsions of the displacement front~(\cite{Drazer04}) may appear and grow linearly with time. Moreover, in highly distorted parts of the front, transverse concentration gradients appear and induce a transverse tracer flux that further enhances dispersion. 

Another important parameter influencing miscible displacements is the rheology of the flowing fluids (relevant, for instance, to enhanced oil recovery using polymer solutions~(\cite{Bird87}). 
Non linear rheological properties modify indeed the flow velocity field  and, more
specifically, the flow velocity contrasts~(\cite{Shah95,Fadili02}).
In the case of shear-thinning fluids in simple geometries like tubes or parallel plates, the flow profile is no longer parabolic but flattens in the center part of the flow channels where the shear rate is lowest :  this decreases the dispersion coefficient (compared to the Newtonian case) but the square law variation of the dispersion coefficient with the Peclet number is still satisfied. For instance, when  the viscosity $\mu$ varies with  the shear rate $\dot{\gamma}$  following a power law :  $\mu \propto \dot{\gamma}^{n-1}$,  Eq.~\ref{eq1} remains valid  but $f$ is a function of $n$~(\cite{Vartuli95}).  

In heterogeneous media, on the contrary, numerical investigations suggest that the flow of shear thinning fluids gets concentrated in a smaller number of preferential flow paths than for Newtonian ones~(\cite{Shah95,Fadili02}): the macrodispersion reflecting large scale distorsions of the displacement front is then increased (even though the local dispersion due to the flow profile in individual channels is still reduced). 
Finally, in this work and in contrast with oil recovery,  the polymer concentration  in the injected and displaced fluids is the same : we investigate only its influence on the transport of a passive solute.
The influence of the shear thinning properties is studied by running experiments with different polymer concentrations.
\section{Experimental set-up and procedure}
\label{sec:sec1}
\subsection{Model fractures and fluid injection set-up}
\begin{figure} [htbp]
\includegraphics[width=\W]{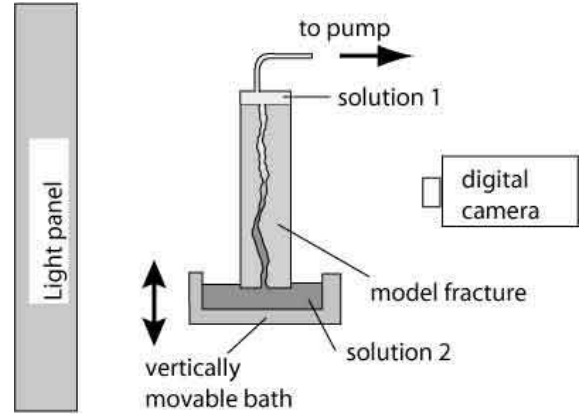}
\caption{Schematic view of the experimental setup. }
\label{fig1}
\end{figure}
Model fractures used in the present work are made of two complementary transparent rough self-affine surfaces clamped together.   
A self-affine surface  is first generated numerically  using the mid-point algorithm (\cite{feder88}) with a self-affine exponent $\zeta$ equal to the value $0.8$ measured for many materials, including granite~(\cite{Bouchaud03}). 
Then, the surfaces are carved by a computer controlled milling machine into a parallelepipedic plexiglas block. The final steps of the machining require an hemispherical tool with a $600\ \mu m$ diameter. The effective size of the surface is  $171$ by $85\ mm$ and the difference in height between the lowest and highest points of the surface is $19.2\ mm$ while the mean square deviation of the height is $3\ mm$.  The two surfaces are exactly complementary but for a $0.33\ mm$ relative shift parallel to their length;  they are bounded on their larger sides by $10\,mm$ wide borders rising slightly above the surfaces. The geometries of these borders is chosen so that they match perfectly and act as spacers leaving an average mean distance  $a_0 = 0.75\ mm$ between the surfaces when the blocks are clamped together. In all cases, the gap between the surfaces is large enough so
that the two walls do not touch : both the mean aperture $a_0$ and the relative displacement are the same in all experiments. 

The fracture assembly is positioned vertically (Fig.~\ref{fig1}) with the two  vertical sides (corresponding to the borders) sealed while the two others are open. The upper side of the model  is connected to a syringe pump sucking the fluids upwards out of the fracture. The lower horizontal side is dipped into a reservoir which may be moved up and down.
The fracture is first saturated by sucking fluid out of the lower reservoir into the model. Then, the original fluid is replaced by the other after lowering the reservoir before raising it again and starting the displacement experiment  by pumping fluid at the top of the model. This procedure avoids unwanted intrusions  while replacing a fluid by the other and allows to purge completely the lower reservoir; a perfectly straight front between the injected and displaced fluids is obtained in this way at the onset of the experiment.
\subsection{Fluid preparation and characteristics}
\label{sec:secfluids}
In all experiments, the injected and displaced fluids are
identical water-scleroglucan solutions but for a small amount of Water Blue dye~(\cite{handbook-dyed}) added to one of the solutions at a concentration
 $c_0 = 0.3 g/l$.  $NaCl$ is added to the other solution to keep both densities  equal. 
 The dye has been chosen such that it has no physico-chemical interaction with the model walls and can be considered as a passive tracer.
The molecular diffusion coefficient of the dye  $D_m \simeq 6.5 \times 10^{-10} \, m^2.s^{-1}$  is determined independently from Taylor dispersion measurements in a capillary tube.

The rheological properties of the scleroglucan solutions have been characterized using a {\it Contraves LS30} Couette rheometer in  range of shear rates  $0.016\,s^{-1}\le \dot{\gamma} \le  87\,s^{-1}$. The rheological properties of the solutions have been verified
to be constant with time within experimental error (over a time lapse of $3$ days) and to be identical  for the dyed and transparent
solutions (for a same polymer concentrations). The variation of the viscosity $\eta$ with $\dot{\gamma}$ is well fitted by the classical  Carreau formula : 
\begin{equation}\label{powervisc}
\eta = \frac{\eta_0 - \eta_{\infty}}{(1+ (\frac{\dot{\gamma}}{\dot{\gamma_0}})^2)^{\frac{1-n}{2}}}  + \eta_{\infty}.
\end{equation}
The values of these rheological parameters for the polymer solutions used in the
present work are listed in Table~\ref{tab1}. $\eta_{\infty}$ is taken equal to the value of the solvent viscosity ($10^{-3}\,Pa.s$ for water) since its determination would require measurements beyond the experimental range limited to  $\dot{\gamma} = 87\,s^{-1}$. In Eq.(\ref{powervisc}), $\dot{\gamma_0}$ corresponds to a crossover between two regimes. For $\dot{\gamma}<\dot{\gamma_0}$, the viscosity $\eta$ tends to $\eta_0$, and the fluid behaves as a newtonian fluid. On the other hand, for $\dot{\gamma}>\dot{\gamma_0}$, $\eta$ decreases following a  power law $\mu \propto {\dot{\gamma}}^{(n-1)}$. Note that, due to the small volume fraction of polymer, the molecular diffusion coefficient keeps the same value as in pure water.
\begin{table}[htbp]
\begin{tabular}{lccc}
Polymer Conc. & $n$ & $\dot{\gamma_0}$ & ${\eta}_0$\\
 ppm          &     & $s^{-1}$       & $mPa.s$\\
$500$ & $0.38 \pm 0.04$ & $0.077 \pm 0.018$ & $410 \pm 33$ \\
$1000$ & $0.26 \pm 0.02$ & $0.026 \pm 0.004 $ & $ 4490 \pm 342$ \\
\end{tabular}
\caption{Rheological parameters of scleroglucan solutions used
in the flow experiments.}
\label{tab1}
\end{table}
\subsection{Optical relative concentration measurements}
The flow rate is kept constant during each experiment and ranges  between $0.01$
 and $1\ ml/min$. The total duration of the experiments  in order to obtain  a complete saturation of the fracture by the invading fluid varies between $20\ min$ and $33\ hours$. 
The transparent fracture is back illuminated by a light panel : about $100$ images of 
the distribution of  light transmitted through the fracture are recorded at constant
intervals during the fluid displacement  using a Roper Coolsnap HQ digital camera with a high, 12 bits, dynamic range. Reference images are recorded both before the experiments and after the full saturation by the displacing fluid in order to have images corresponding to the fracture fully saturated with both the transparent and the dyed fluid. 

\begin{figure}[h]
\includegraphics[width=\W]{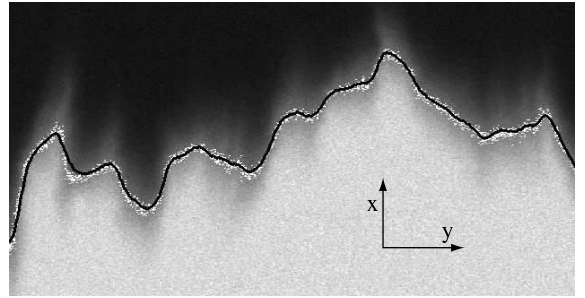} 
\caption{Experimental relative concentration field obtained with a $1000\ ppm$ polymer concentration for a mean front displacement of half the fracture length. Grey levels represent values of the ratio  $c(x,y,t)/c_0$. Size of field of view : $81\ mm \times 70\ mm$;  only a part of the actual image is shown. Solid line : front profile $x_f(y,t)$ as defined in section~\ref{macrostructure}; white dots are pixels where $c(x,y,t)/c_0\, = \,0.5 \pm 0.03$.}
\label{fig2}
\end{figure}

The local relative concentration of the displacing fluid (averaged over the fracture aperture) is determined from these images by the following procedure.
First,  the absorbance $A(x,y,t)$ of light by the dye on an image obtained at the time $t$ is computed by the relation :
 \begin{equation}
A(x,y,t) = ln(\frac{I_t(x,y)}{I(x,y,t)})
\label{eq:calib}
\end{equation}
in which $I_t(x,y)$ and $I(x,y,t)$ are the transmitted light intensities (in grey levels) measured for a pixel of coordinates $(x,y)$ respectively when the fracture is saturated with transparent fluid and at time $t$.
When the fracture is saturated with the dyed fluid ($ c(x,y) = c_0$), the transmitted light intensity is  $I_0(x,y)$ so that  the adsorbance $A_0(x,y)$  is : 
\begin{equation}
A_0(x,y) = ln(\frac{I_t(x,y)}{I_0(x,y)})
\label{eq:calib0}
\end{equation}
The relation between the local concentration $c(x,y,t)$ (averaged  over the local fracture aperture), the dye concentration $c_0$ in the dyed fluid and the absorbances $A$ and $A_0$ has been determined independently from calibration pictures realized with 
the fracture saturated with dyed solutions of concentrations $c$ increasing from $0.1$ to $0.5\ g/l$.
The ratio $A(x,y)/A_0(x,y)$ is found experimentally  to be constant  within $\pm\,3\%$ over the picture area  :  for more precision, the ratio $<A>_{x,y}/<A_0>_{x,y}$ of the averages  is therefore used to determine the calibration curve. Due to non linear adsorbance~(\cite{Detwiler00}),  the relation $c/c_0 = <A>_{x,y}/<A_0>_{x,y}$  predicted by Beer-Lambert's law is not valid. The variation of $c/c_0$ with $<A>/<A_0>$ follows however accurately the polynomial relation :
\begin{equation}
\frac{c}{c_0}=b_1 \frac{A}{A_0} + b_2 (\frac{A}{A_0})^2 + b_3 (\frac{A}{A_0})^3
\label{eq:eq}
\end{equation}
 with $b_1 = 0.186\pm0.023$, $b_2=0.0087\pm 0.04$ and $b_3=0.108\pm 0.021$.
 Practically, Eq.~\ref{eq:eq} is applied to all pixels $(x,y)$ in the pictures recorded during the experiment in order to obtain $c(x,y,t)/c_0$. An instantaneous relative concentration map obtained in this way  is displayed in Fig.~\ref{fig2}. 
In the following, $c_0$ is omitted and c(x,y,t) refers to the local {\it relative} concentration at a given time (still averaged as above over the fracture aperture).
\section{Local concentration variations}
As already pointed above, transport in the fracture results from the combination of front spreading due to large scale flow velocity variations and of mixing due to local dispersion mechanisms and concentration gradients.  In order to identify these different processes, a local analysis is first performed. 
\begin{figure}[h]
\includegraphics[width=\W]{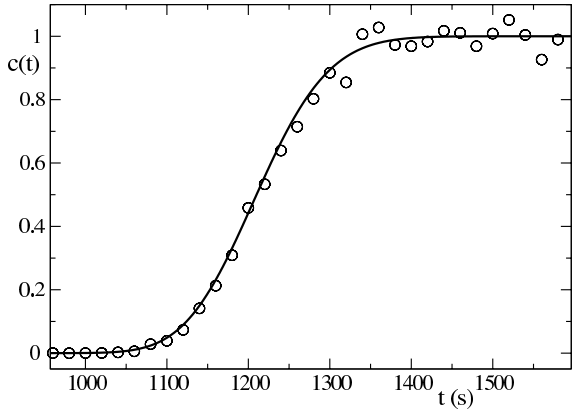} 
\caption{Time variation of the relative concentration $c(x,y,t)$ for $x=20\, mm$ and $y=36\, mm$ for $1000\, ppm$ polymer solutions. Mean flow velocity $U=0.014 mm/s$ ($Pe = 150$). Solid line : fit by Eq.\ref{eq:eq3} with $\overline{t(x,y)}=1212\, s$ and $D(x,y) = 0.315\, mm^2/s$.}
\label{fig3}
\end{figure}
For  each pixel $(x,y)$, the variation $c(x,y,t)$ with time of the local relative concentration of the dyed fluid has been determined; as can be seen in Fig.~\ref{fig3}, it is well fitted by the following  solution of the convection-dispersion equation corresponding to the stepwise concentration variations induced experimentally :
\begin{equation}
c(x,y) = \frac{1}{2} (1 + erf \frac{t-\overline{t(x,y)}}{\sqrt{4\frac{D(x,y)}{U^2} t}})
\label{eq:eq3}
\end{equation}
In Eq.~\ref{eq:eq3}, $U$ is the mean velocity, $D(x,y)$ is the local dispersion coefficient and $\overline{t(x,y)}$ the mean transit time. Note that, if the injected fluid  is the transparent one, the + sign should be replaced by a - in Eq.~\ref{eq:eq3}. Both $D(x,y)$ and $\overline{t(x,y)}$ are defined locally and may depend on the measurement point. 

In the following, the variations of these quantities are analyzed : on the one hand, the spatial variations of $\overline{t(x,y)}$ reveal the channelized structure of the flow that leads to macrodispersion. On the other hand, $D(x,y)$ reflects local mixing processes taking place on each flow line.   
\section{Local dispersive mixing}
\label{sec:local}
\begin{figure}[htbp] 
\includegraphics[width=\W]{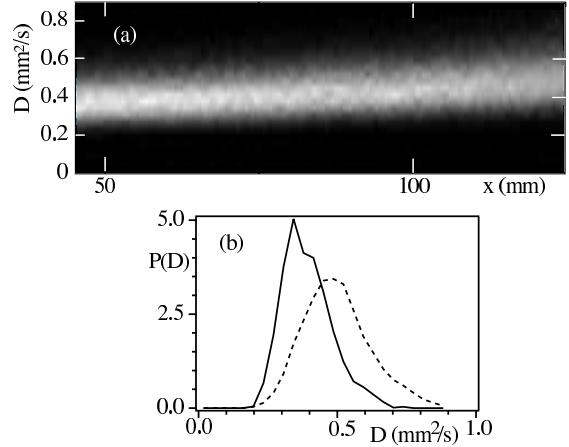} 
\caption{Probability distribution $P(D,x)$ of the dispersion coefficient for a dyed solution displacing a transparent one : mean flow velocity $U=0.014\, mm/s$ ($Pe = 150$), polymer concentration  $1000\ ppm$). - (a) Grey levels correspond to the value of the probability P(D) at the corresponding values of $D$ and $x$  - (b) Distributions $P(D)$ respectively for $x = 45$ (solid line) and $125 \, mm$ (dashed line).}
\label{fig4}
\end{figure}

In this section, the variation of $D(x,y)$ is studied as a function of the distance $x$ from the inlet  and of the fluid velocity and rheology. The probability distribution of the local values of $D(x,y)$ determined for all pixels at a same distance $x$ for a given experiment is displayed in Fig. \ref{fig4}a in a grey level scale as a function of  $x$  ($45 \le x \le 125 mm$); Fig. \ref{fig4}b shows the two extreme distributions corresponding to $x = 45$ (solid line) and $125 \, mm$.  
One observes that the mean value $\overline D$ of the distributions increases only  by $20 \%$  between  $x = 45$ and $125 \, mm$. $x$ ($\overline D$ will just be refered to as $D$ in the following and  the deviations of the local values  will be characterized by the width $\Delta D$ 
of $P(D)$ at mid-height  which increases also slowly with distance). The drift of $D$ may be due to slow variations of the mean aperture and flow velocity.  It may also reflect an increased dispersion   in distorted regions of the front  : there, dye  diffuses across the flow lines which  contributes to  broaden the front.
The slow variation of the mean value of  $D$  with $x$  together with the good fit displayed  in Fig.~\ref{fig3}, demonstrate that the Fickian dispersion model describes well  the local spreading of the front during all the experiment . 
The symmetry of the process is finally checked by realizing the same experiments with the transparent  solution displacing the dyed one. The distributions of the dispersion coefficients for given values of $x$ are the same as in the reverse configuration : there is therefore no effect of small residual density contrasts..

\begin{figure}  [htbp]
\includegraphics[width=\W]{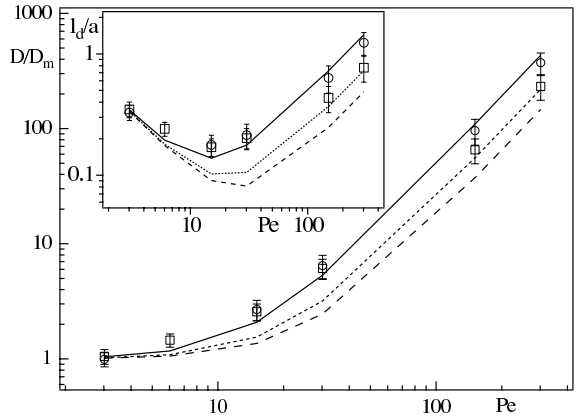}
\caption{Variation of normalized dispersion coefficient $D/D_m$ with $Pe = Ua_0/D_m$. Insert : variation of  the dispersivity $l_d=D/U$. Solid, dotted and dashed lines : predictions from Eq.~\ref{eq1} respectively for $n=1$ (Newtonian fluid), $n=0.38$ and $n=0.26$ (shear thinning solutions with respective $500\, ppm$ and $1000\, ppm$ polymer concentrations).}
\label{fig5}
\end{figure}

The dependence of the local dispersion on the flow velocity $U$ for the $500 \, ppm$ and $1000 \, ppm$ solutions is displayed in  figure~\ref{fig5}  where $D$ is plotted as a function of the P{\'e}clet number $Pe$ defined above. The same variation trends are followed for both solutions :  for low $Pe$ values the dispersion coefficient tends towards a constant close to $1$, while at high $Pe$, $D$ increases as the square of $Pe$. These variations are similar to the predictions of Eq.~\ref{eq1} (solid, dotted and dashed lines) also plotted in figure~\ref{fig5} for a newtonian fluid and two power law shear thinning fluids for which $\mu \propto \dot{\gamma}^{n-1}$ ($n$ is taken equal to the values listed in Tab.~\ref{tab1} and the value of $f$ in Eq.\ref{eq1} is computed from Eq.~$4$ in \cite{Boschan03}). 
The overall agreement observed implies that Taylor dispersion is indeed the dominant mechanism controlling local dispersion. 

In a more detailed analysis, one must however  take into account the fact that, for real fluids, the viscosity does not diverge at low shear rates but becomes constant (Newtonian plateau viscosity) for ${\dot \gamma}<\dot{\gamma_0}$. In a Poiseuille Newtonian flow between two parallel flat plates, the shear rate is maximum at the wall with $\dot\gamma = 6U/a$. It follows that the transition value $\dot{\gamma_0}$ is reached for $U = U_c = a \dot{\gamma_0}/6$. 
Using the values of $\dot{\gamma_0}$ in Tab.\ref{tab1}, the velocities $U_c$ (resp. P\'eclet numbers $Pe_c$) corresponding to the $500\, ppm$ and $1000 \, ppm$ solutions are $0.01\, mm/s$ and $0.003\, mm/s$  (resp. $Pe_c\,=\,11$ and $4$). Below $Pe_c$, the dispersion coefficient should be the same as for a Newtonian fluid; above $Pe_c$, its variation should progressively merge with that predicted for power law fluids.
This crossover is clearly observed in Fig.\ref{fig5} : for $Pe\,<\,30$, values of $D$ obtained with the $500\, ppm$ and $1000\, ppm$ solutions  co\"\i ncide with the predictions for Newtonian fluids. At high P\'eclet numbers, data points corresponding to the two solutions get separated and the values of $D$ become close to the  predictions for power law fluids with the corresponding values of $n$. 

It is often convenient to replace the dispersion coefficient $D$ by the dispersivity $l_d = D/U$ to identify more easily the influence of spatial heterogeneities of the flow field. For Taylor like dispersion verifying Eq.\ref{eq1} and for a Newtonian fluid, the dispersivity has a minimum equal to $l_d=2a/\sqrt{210} \approx 0.0095 a$ for $Pe=\sqrt{210} \simeq 14.5$. The insert of Fig.\ref{fig5} displays the variation of the dispersivity with $Pe$ :  its minimum value corresponds well to the theoretical prediction for Newtonian fluids (solid line). This confirms that,  in this range of P{\'e}clet numbers, the two polymer solutions  behave like Newtonian fluids.

These results demonstrate that the local dispersion $D(x,y)$ in the rough model fracture is mainly due to the flow profile in the gap between the walls and similar to that between flat parallel plates. This is likely due to two properties of the flow field : first, as for parallel plates, flow lines initially located in the center of the gap remain there during their full path through the fracture and, similarly, those located close to a wall only move away from it through molecular diffusion. Also,  the orientation of the local flow velocity is always close to that of the mean flow as will be seen below.
In the next section, we discuss on the contrary macrodispersion due to  variations in the plane of the fracture of the local velocities (averaged this time over the gap).
\section{Macrodispersion in the model fractures}\label{sec:global}
\subsection{Flow structure and mean front profile}\label{subsec:global}
\label{macrostructure}
As pointed above,  the complementary rough walls of the fracture model are translated  relative to each other in the direction $y$ perpendicular to the mean flow (along the $x$ axis) : in this configuration, large scale channels parallel to $x$ appear~(\cite{GentierLAR97,Drazer04,Auradou05}) with only weak variations of the flow velocity along their length~(\cite{Auradou06}). In the following, we use  a simple model in which the fracture is described as a set of independent parallel channels where the effective flow velocity $U(y)$ depends only on the transverse coordinate.
\begin{figure} [htbp]
\includegraphics[width=\W]{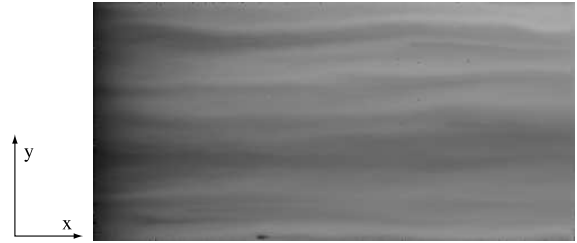}
\caption{Greyscale map of the normalized local transit time $\overline{t(x,y)}U/x$. Flow is from left to right. Dark (resp. light) regions correspond to locations where $\overline{t(x,y)}U/x <1$ (resp. $\overline{t(x,y)}U/x>1$).  Mean flow velocity  $U = 0.014\,mm/s$  ($Pe = 150$), polymer concentration $1000\, ppm$).}
\label{fig6}
\end{figure}

 The validity of this assumption is tested in Fig.~\ref{fig6} in which the values of the normalized transit time $\overline{t(x,y)}U/x$ are represented as grey levels at all points  $(x,y)$ inside the field of view; $\overline{t(x,y)}$ is the local effective transit time determined by fitting the curve of Fig.~\ref{fig3} corresponding to point $(x,y)$ by solutions of Eq.~\ref{eq:eq3}.  
Dark (resp. light) pixels mark points where  $\overline {t(x,y)}U/x$ is respectively higher (resp. lower) than $1$ : dark and light streaks globally parallel to $x$ are clearly visible and extend over the full length of the model fractures. These streaks correspond to slow (resp.  fast) flow paths and  their orientation  deviates only slightly from $x$ : this is in agreement with the above simple model of parallel flow paths with different velocities remaining correlated along the full path length.
 
Another important feature of the maps  of Fig.~\ref{fig6} is that they allow to determine an instantaneous front profile  $x_f(y,t)$ at a given time $t$ :  in the following, it will be defined as the set of all points for which
 ${\overline t(x,y)} \, = \, t$. In our experiments this profile was  very similar to the  isoconcentration line  $c(x,y,t)/c_0 \, =\, 0.5$ (as can be seen  in Fig.~\ref{fig2}). In the following, the macrodispersion process  will be directly characterized by the variations of these  front profiles with time without analyzing extensively the concentration maps. 
\subsection{ Global front dynamics}
\label{subsec:frontdyn}
\begin{figure} [htbp]
\includegraphics[width=\W]{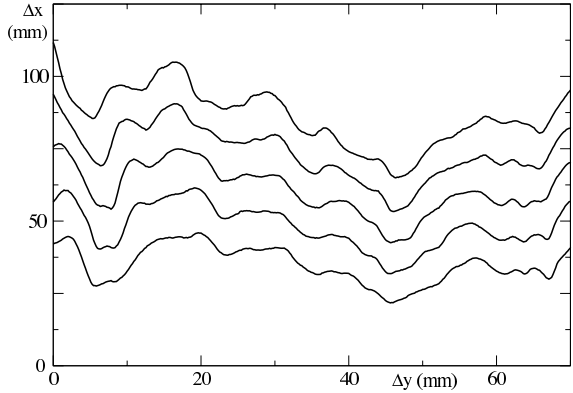}
\caption{Front profiles $x_f(y,t)$ at times $t=6.5$, $8$, $9.5$, $11$ and $12.5 \, min$ for a $1000\, ppm$ polymer solution and $U = 0.014\,mm/s$  ($Pe = 150$) The mean flow velocity is oriented from the bottom to the top of the figure).}
\label{fig7}
\end{figure}

Fig.\ref{fig7} displays several  front  profiles obtained at different times by this procedure. As expected, the profile is initially quite flat but large structures appear and grow with time.  A key feature is the fact that similar structures are observed on all fronts : it is only their size parallel to the mean flow that increases with time. This confirms the large correlation length  parallel to $x$   of the high and low velocity regions  and, therefore, the flow channelization already assumed above.
 
\begin{figure} [htbp]
\includegraphics[width=\W]{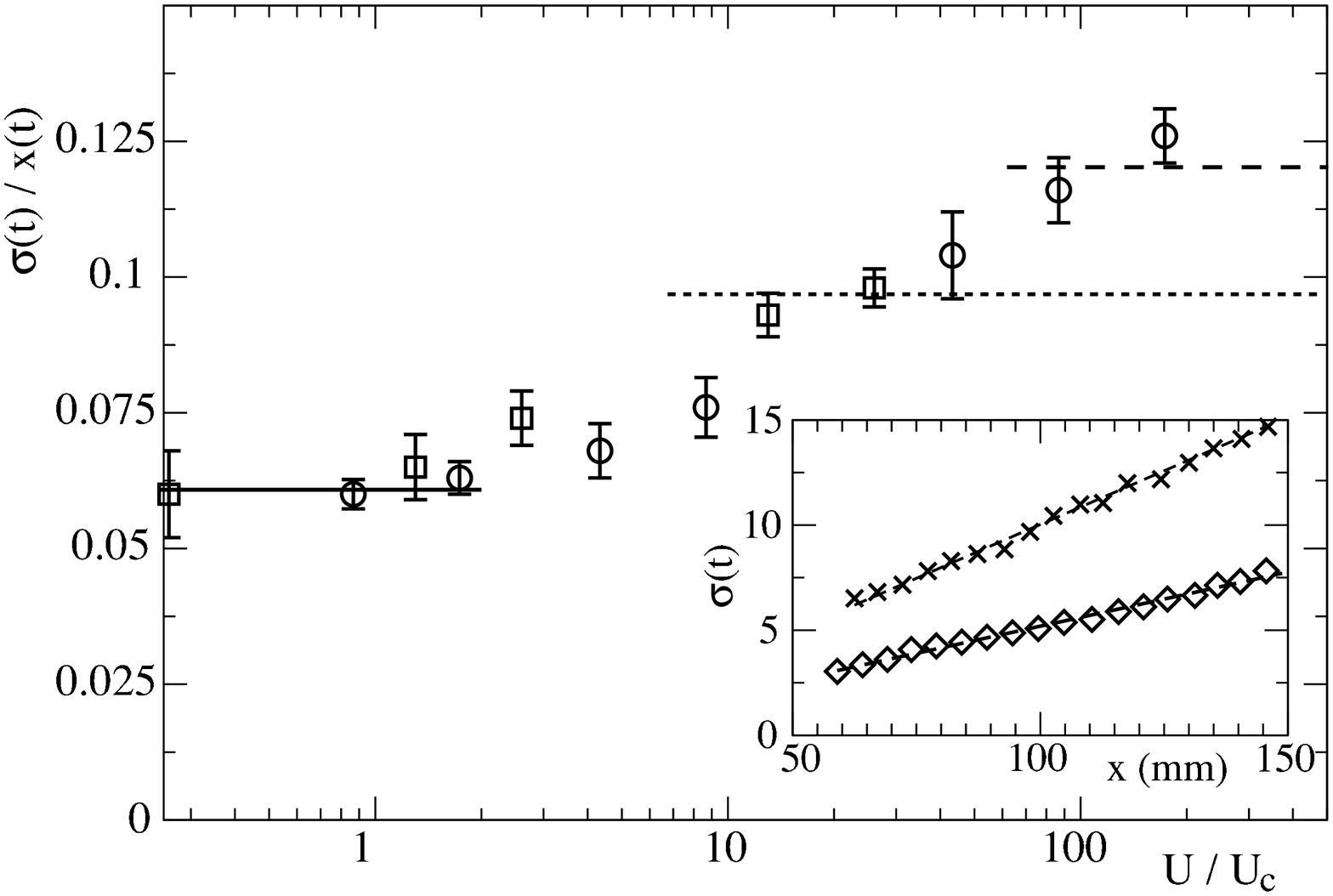}
\caption{Variation in Log linear coordinates of the relative front width $\sigma (t) / \overline{x(t)}$ with the normalized mean velocity $U/U_c$;  crossover velocity $U_c=0.01\, mm/s$ (resp. $0.003\, mm/s)$ for ($\square$) : $500\, ppm$  (resp. ($\circ$) $1000\, ppm$) solutions. Solid, dotted and dashed horizontal lines : predicted values respectively for a Newtonian fluid and power law fluids with same index $n$ as the  $500\, ppm$ and $1000\, ppm$ solutions. Insert :  variation of $\sigma (t)$ ($mm$) with the distance $x$ for a $1000\, ppm$ solution. Diamonds :$U =0.0056\,mm/s$  ($Pe = 60$), Crosses :$U =0.056\,mm/s$ ($Pe = 600$). Dashed line : linear regression.}
\label{fig8}
\end{figure}

In such cases, as pointed out by \cite{Drazer04}, the size of the structures of the front, and therefore its global  width $\sigma(t)$ should increase linearly with distance (and time).  In the following, the width  $\sigma(t)$ is defined by  $\sigma^2(t) = <(x_f(y,t) \,-\,<x_f(y,t)>_y)^2>_y$ in which $<x_f(y,t)>_y$ is the mean distance  of the front from the inlet at time t : as shown in the insert of Fig.\ref{fig8}, $\sigma(t)$ increases as expected linearly with time. 
Since the mean distance $<x_f(y,t)>_y$ also increases linearly with time, the ratio $\sigma(t)/<x_f(y,t)>_y$ remains constant and may therefore be  used to characterize the magnitude of the macrodispersion. In the framework of the simple model assuming parallel independent channels with different flow velocities, 
 $\sigma(t)$ and $<x_f(y,t)>$ are respectively of the order of $\delta U \, t$ and $Ut$ where $\delta U$ the root mean square of the velocity variations between the different channels. Therefore, the ratio  $\sigma(t)/<x_f(y,t)>_y$ corresponds to the relative magnitude  $\delta U / U$ of the velocity contrasts inside the fracture (a well known result for stratified media with negligible exchange between layers).

For both Newtonian ($\eta$ = cst.) and power law ($\eta \propto \dot{\gamma}^{n-1}$) fluids  the relative velocity fluctuations $\delta U \propto U$ are expected to be constant with $U$ :  $\sigma(t)/\overline{x(t)}$ should therefore be independent of the flow rate $Q$. The variations of  $\sigma(t)/\overline{x(t)}$ with the normalized velocity $U/U_c$ for the two polymer solutions used in our experiments  are displayed in Fig.\ref{fig8} ($U_c$ is the  cross-over velocity  between the Newtonian and shear thinning behaviours introduced in section.~\ref{sec:local}), The values of $\sigma (t) / \overline{x(t)}$ are averages over several time intervals (error bars indicate fluctuations with time). For $U<U_c$,  $\sigma(t)/\overline{x(t)}$ retains a constant value close to $0.05$ independent of the polymer concentration which likely corresponds to a Newtonian behaviour. For  $U>U_c$, $\sigma(t)/\overline{x(t)}$ increases faster with $U$  for the more concentrated ($1000 \ ppm$) solution.  The normalized widths should reach a new limit at higher flow rates  ($U\gg U_c$) : theoretical values for the two solutions are indicated as horizontal lines.

More precisely, the shear rate $\dot{ \gamma}$  is always zero in the middle of the gap of the fracture and highest at the walls. If the shear rate at the wall is larger than the transition value $\dot{\gamma}_0$ (see section~\ref{sec:local}), there are two domains in the velocity profile :  the fluid rheology is   Newtonian in the central part of the fracture and non Newtonian near the walls~(\cite{Gabbanelli05}). When the flow rate Q increases, the fraction of the flow section where flow is Newtonian shrinks  while the fraction where it is non Newtonian expands : for a shear thinning fluid with a power law characteristic ($n<1$), the average velocity in a given flow channel (the  integral of the velocity profile over the fracture gap) increases faster with the longitudinal pressure gradient (as $\nabla p^{1/n}$) than  for Newtonian fluids (as $\nabla p$). This enhances velocity contrasts between channels of different apertures (assuming that they are subject to similar pressure gradients)  and finally increases the front width  compared to the Newtonian case. These effects get larger as the concentration increases from $500$ to $1000\,ppm$ both because the exponent $n$ decreases and because the transition shear rate  $\dot{\gamma}_c$ is smaller (Tab.~\ref{tab1}). This lower value of 
$\dot{\gamma}_c$  for  the $1000\,ppm$ solution does indeed increas  the fraction of the flow section where the fluid rheology is  shear thinning.
\subsection{Front geometry}
\begin{figure} [htbp]
\includegraphics[width=\W]{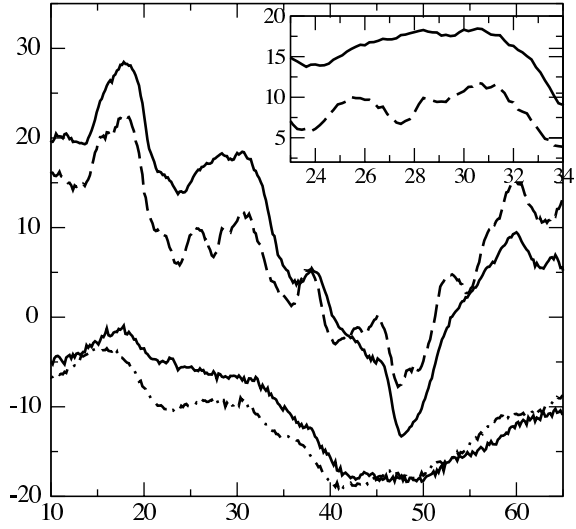}
\caption{Front geometries for a mean distance from the inlet equal to  half the fracture length. Solid (resp. dashed) lines : $1000 \, ppm$ (resp. \,$500 ppm$) solutions. Top (resp. bottom) curves : $U=0.3\, mm/s$ i.e.  $Pe = 3000$ (resp. $U=0.003\, mm/s$ i.e.  $Pe = 30$). The two sets of curves have been shifted to allow for easier comparisons. Insert : close up of the upper curves ($U=0.3\, mm/s$ i.e.  $Pe = 3000$).}
\label{fig9}
\end{figure}
In the previous section, the overall front width has been shown to depend on the global flow rate and on the fluid rheology;  their influence on the detailed front structure will now be analyzed. Figure~\ref{fig9} compares front geometries observed for the two solutions used in the experiments at the lowest  (resp. highest) mean velocities investigated : $U = 0.003\,mm/s$ (resp. $0.3\,mm/s$).  The lower velocity is below $U_c$ and the rheology of both fluids is therefore Newtonian. The  polymer concentration plays then a minor part and the front geometries are very similar (lower curves). In addition, at this mean velocity, the P{\'e}clet number is $\approx 3$ and therefore lower than the value  $Pe \simeq 14.5$ (see section~\ref{sec:local}) corresponding to the crossover between Taylor dispersion and longitudinal molecular diffusion. As a result, transport at the local scale is controlled by molecular diffusion which smears out the effect of local velocity fluctuations and smoothens the front geometry.

The higher  velocity  $U=0.3\, mm/s$ (upper curves) is well above $U_c$ :   the global  front width  increases then with the polymer concentration (see section~\ref{subsec:frontdyn}), as  can be seen for the two upper curves in  Fig.\ref{fig9}.  However, at smaller length scales,  the geometrical characteristics of the front  and their dependence on the fluid rheology are more complex. Geometrical features of lateral size below $10\, mm$ have, for instance, a larger extension parallel to the mean flow for the $500\, ppm$ solution.

\begin{figure} [htbp]
\includegraphics[width=\W]{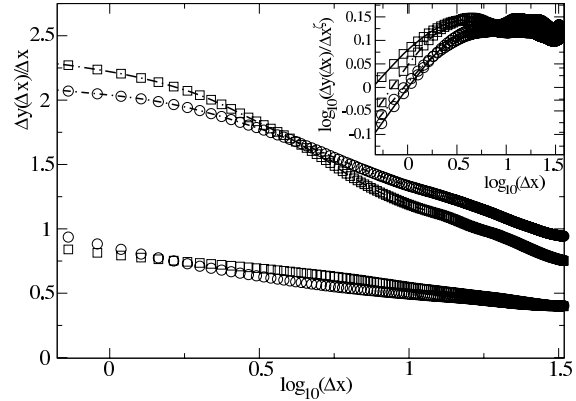}
\caption{Variation of the ratio $\Delta y(\Delta x)/\Delta x$ as a function of $log_{10}(\Delta x)$ for the same fronts as in Fig.~\ref{fig9} with $\circ$ (resp. $\square$) :  $1000\, ppm$ (resp. $500\, ppm$) polymer  solutions. Symbols with no lines : $U=0.003\, mm/s$  ($Pe = 30$) and with dotted lines : $U=0.3\, mm/s$ ($Pe = 3000$). Insert : variation of  $log_{10} (\Delta y/\Delta x^{\zeta_f})$ as a function of $\log_{10} (\Delta_x)$ ($\zeta_f$ has been chosen equal to $0.8$ in order to provide the best fit with a power law at large  $\Delta x$ values).}
\label{fig10}
\end{figure}

Previous theoretical,experimental and numerical studies of displacement fronts between sheared complementary self-affine walls indicate that  their geometry is also self-affine (over a finite range of length scales)  with the same characteristic exponent $\zeta$ as the fracture walls (\cite{Roux98,Drazer04,Auradou01}). 
Such self-affine profiles  $y(x)$ may be characterized quantitatively from the maximum variation   $y_{max} - y_{min}$  of $y(x)$ in a window of width 
$\Delta x$ ($y_{max}$ and $y_{min}$ are the maximum and minimum values of $y(x)$ in this window). In  this "min-max" method,  the average $\Delta y(\Delta x)$ of the values of $y_{max} - y_{min}$ is computed for all locations of the window inside the profile and the process is repeated for the different values of  $\Delta x$. For a self-affine curve of characteristic roughness exponent $\zeta_f$, one has, for instance,  $\Delta y \propto \Delta x ^\zeta_f$ ($\zeta_f = 1$ corresponds to a Euclidian curve).

This result is verified in the insert of  Fig.\ref{fig10} where  the ratio $\Delta y/\Delta x^{\zeta_f}$ is plotted as a function of $\Delta x$ in log-log coordinates for $\zeta_f = 0.8$ : $\Delta y/\Delta x^{\zeta_f}$ is indeed observed to remain constant over a  broad range of variation of $\Delta x$ (see insert of Fig.~\ref{fig10}).  The lower boundary of this self affine domain  increases slightly with the polymer concenration from $\approx 3\, mm\, (500\, ppm)$ to $\approx 5\, mm\,(1000 \, ppm)$ and depends weakly on the flow velocity (provided $Pe\gg 1$). For 
$\Delta x$ below this cross over length, the slope of the curves is close to $0.2$, reflecting an Euclidean geometry with $\Delta y \propto \Delta x$. 

In order to analyse the influence of the fluid properties and of the flow velocity $U$ on these results, the variation of  the ratio $\Delta y/\Delta x$ with $\Delta x$ is displayed in Fig.~\ref{fig10} for the two polymer solutions and for two different values of $U$ ($U<U_c$ and $U>U_c$).  $\Delta y$ has been normalised by the window width $\Delta x$ to reduce the amplitude of the global variations of $\Delta y$ and make more visible the differences between the curves. 

For the lowest velocity $U$, the curves are similar for both polymer concentrations (as expected for a Newtonian rheology). The ratio $\Delta y/\Delta x$ increases with $U$,  reflecting a higher amplitude of the geometrical features of the front at all length scales : in addition to this global trend,  the variation of $\Delta y$ with $\Delta x$ depends however significantly on  the polymer concentration.  
For  $U>U_c$ (upper curves in Fig.~\ref{fig10}),  features of the front with large transverse sizes $\Delta x$ are of larger amplitude $\Delta y$ for the more concentrated solution (as noted above); on the contrary, smaller features corresponding to low $\Delta x$ values are more developed for the less concentrated solution (this is qualitatively visible on Fig.~\ref{fig9}). The two curves cross each other  for $\Delta x \approx 4\, mm$. This attenuation of small scale features of the front for the more concentrated solution  may reflect an enhancement of the transverse diffusion of the fluid momentum due to its higher viscosity (in other words, drag forces between parallel layers of fluid moving at different velocities become larger). This smoothens out both the local velocity gradients and the associated small features of the front but does not influence large scale velocity variations. These are due to effective aperture contrasts between between parallel channels : their influence is amplified when the exponent $n$ decreases for higher polymer concentrations, resulting in a larger global front width parallel to the flow.

\section{Discussion and conclusion}
Studying miscible displacement processes by an optical method in a transparent model fracture  has revealed important characteristics of  flow and transport in rough fractures : these results may be applicable to fluid displacements and transport  in fractured reservoirs. A major feature of  this approach is the possible simultaneous analysis of both local mixing and global front spreading  due to large scale heterogeneities :  the different transport mechanisms may in particular be characterized by maps of the local  transit time from the inlet and of the local dispersion coefficient.

The multiple length scales features of natural fractures have been reproduced  by assuming  rough walls of complementary  self-affine geometries and with a relative displacement parallel to their mean plane (this models the effect of shear during fracturing). In the present experiments, this relative displacement was perpendicular to the mean flow : this induced a channelization of the flow field with small velocity variations along the flow lines and larger ones across them. 
Large macrodispersion effects are observed in such a geometry : in the present experiments, the front width increases linearly with distance from the inlet and its structure reflects closely the velocity variations between the different parallel channels. 

In contrast,  front spreading at the local scale remains diffusive : moreover, the corresponding dispersion coefficient is close to that estimated for Taylor dispersion in an Hele-Shaw cell with plane walls separated by a distance of the order of the main aperture of the fracture. This value may however be locally increased by transverse diffusion in highly distorted regions of the fronts where they  get locally parallel to the mean flow.

These fluid displacements are strongly influenced by the rheology of the flowing fluids  so that additional informations can be obtained by varying the polymer concentration and/or the shear rate.
At low shear-rates (below a transition value $\dot \gamma_0$), the polymer solutions used in the present work behave like Newtonian fluids and their concentration has no effect at low mean velocities ($U < U_c$). At higher shear rates ($U > U_c$), the shear thinning effects become significant : they increase with the polymer concentration but may be very different depending on the scale of observation. The global width $\Delta x$ of the front parallel to the mean flow gets larger at larger polymer concentrations while, in contrast, smaller geometrical features of the front  are reduced. In addition, for $U>U_c$, polymer also influence at the local scale the Taylor-like dispersion due to the flattening of the flow profile between the fracture walls.
Such rheological effects may strongly influence the efficiency of enhanced oil and waste recovery processes.

Such results  raise a number of questions to be answered in future work. First, one may expect the spatial correlations of the velocity to decay with distance, leading finally to  normal Fickian dispersion.  The present samples were not long enough to allow for the observation of this transition : it may however be more easily observable for models designed so that flow is parallel to the relative shear of the complementary fracture surfaces (in this case, the correlation length should be smaller).  Another important issue is the influence of contact area on the transport process : one may expect in this case  the development of low velocity regions leading to anomalous dispersion curves.     

\begin{acknowledgments}
We are indebted to G. Chauvin and  R. Pidoux for their assistance in the realization of the experimental
set-up and to C. Allain for her cooperation and advice. HA and JPH are funded  by the EHDRA (European Hot Dry Rock Association)  in the frame work of the STREP Pilot plan program SES6-CT-2003-502706) and by the CNRS-PNRH program.
This work was also supported by a CNRS-Conicet Collaborative Research Grant (PICS $n^{\circ} \, 2178$) and by the ECOS Sud program  $n^{\circ}\,A03E02$.
\end{acknowledgments}
\bibliography{articles,book,mios,mybooks}
\bibliographystyle{agu04}

\end{article}
\newpage

\end{document}